\documentclass{JHEP3}
\usepackage{enumerate,amssymb}
\renewcommand{\a}{\alpha}
\renewcommand{\b}{\beta}
\newcommand{\g}{\gamma}           
\renewcommand{\d}{\delta}         
\newcommand{\e}{\varepsilon}

\newcommand{\ki}{\chi}
\newcommand{\la}{\lambda}

\newcommand{\om}{\omega}         \newcommand{\OM}{\Omega}
\newcommand{\p}{\psi}             

\newcommand{\s}{\sigma}           
         
\newcommand{\f}{{\phi}}           \newcommand{\F}{{\Phi}}

\newcommand{\x}{\xi}              

\newcommand{\z}{\zeta}

\newcommand{\eps}{{\epsilon}}

\newcommand{\ca}{{\cal A}}
\newcommand{\cb}{{\cal B}}
\newcommand{\cc}{{\cal C}}
\newcommand{\cd}{{\cal D}}

\newcommand{\cf}{{\cal F}}
\newcommand{\cg}{{\cal G}}

\newcommand{\ck}{{\cal K}}

\newcommand{\be}{\begin{equation}}
\newcommand{\ee}{\end{equation}}
\newcommand{\eqn}[1]{\label{#1}\end{equation}}
\newcommand{\equ}[1]{(\ref{#1})}

\newcommand{\bea}{\begin{eqnarray}}
\newcommand{\eea}{\end{eqnarray}}
\newcommand{\eqan}[1]{\label{#1}\end{eqnarray}}

\newcommand{\ba}{\begin{array}}
\newcommand{\ea}{\end{array}}

\newcommand{\nn}{\nonumber}

\newcommand{\loco}{{\mathop{ \, \rule[-.06in]{.2mm}{3.8mm}\,}}}
\newcommand{\doubar}{{{\loco}\!{\loco}}}

\newcommand{\az}{{\bf{z}}}
\newcommand{\au}{{\bf{u}}}
\newcommand{\av}{{\bf{v}}}

\newcommand{\da}{{\dot{\alpha}}}
\newcommand{\db}{{\dot{\beta}}}
\newcommand{\dg}{{\dot{\gamma}}}
\newcommand{\dd}{{\dot{\delta}}}

\newcommand{\ta}{{\mbox{\tiny{A}}}}
\newcommand{\tb}{{\mbox{\tiny{B}}}}
\newcommand{\tc}{{\mbox{\tiny{C}}}}
\newcommand{\td}{{\mbox{\tiny{D}}}}
\newcommand{\te}{{\mbox{\tiny{E}}}}
\newcommand{\tf}{{\mbox{\tiny{F}}}}

\renewcommand{\gg}{{\mathfrak{g}}}
                         \newcommand{\hh}{{\mathfrak{h}}}
                         \renewcommand{\tt}{{\mathfrak{t}}}
\newcommand{\xx}{{\mathfrak{s}}}
\newcommand{\qq}{{\mathfrak{q}}}

\newcommand{\vd}{\textrm{d}}
\newcommand{\vi}{\textrm{i}}

\newcommand{\bs}{\bar{\s}}
\newcommand{\bp}{\bar{\p}}

\newcommand{\brg}[1]{\left(#1\right)_T}
\newcommand{\br}[1]{\left(#1\right)_H}

\newcommand{\ym}{{\textrm{\tiny YM}}}
\newcommand{\sg}{{\textrm{\tiny SG}}}
\newcommand{\wz}{{\textrm{\tiny wz}}}
\newcommand{\oo}{{\textrm{\tiny (0)}}}

\title{$N=4$ central charge superspace at work
for supergravity coupled to an arbitrary number of abelian vector multiplets}

\author{S\'ebastien GURRIERI\\
Yukawa Institute for Theoretical Physics\\
Kyoto University\\
Kyoto, 606-8502, Japan\\
E-mail: gurrieri@yukawa.kyoto-u.ac.jp
}

\author{Annam\'aria KISS\\
Center for String and Particle Theory\\
Department of Physics, University of Maryland\\
College Park, MD 20742-4111 USA\\
E-mail: akiss@physics.umd.edu
}

\abstract{We present the description in central charge superspace of
$N=4$ supergravity with antisymmetric tensor coupled
to an arbitrary number of abelian vector multiplets.
All the gauge vectors of the coupled system are treated
on the same footing as gauge fields corresponding to translations
along additional bosonic coordinates. It is the geometry of
the antisymmetric tensor which singles out which combinations of
these vectors belong to the supergravity multiplet and which are the
additional coupled ones. Moreover, basic properties of
Chapline-Manton coupling mechanism, as well as the
$\frac{SO(6,n)}{SO(6)\times SO(n)}$ sigma model of the Yang-Mills
scalars are found as arising from superspace geometry.}


\preprint{}

\begin{document}

\today

\section{Introduction}

Interest in $N=4$ supergravity originates from the fact that it
has the largest amount of supersymmetry that still allows
existence of matter multiplets. The supergravity multiplet
contains one graviton, 4 gravitini, 6 vectors, 4 spin $1/2$
fermions, 1 scalar and 1 pseudoscalar, or, equivalently with
dualization, one scalar and one antisymmetric 2-tensor. Matter
comes only under the form of vector multiplets which contain 1
vector, 4 spinors, and 6 scalars.

$N=4$ supergravity exists in two versions, one with global $SO(4)$
\cite{Das77,CS77b} and one with global $SU(4)$ symmetries
\cite{CSF78}. As long as these symmetries are not gauged, the two
versions are equivalent \cite{CSF78} and we will be interested in
the SU(4) theory whose advantage is to restrict non-polynomiality
to exponentials of the scalar. The Lagrangian and supersymmetry
transformations for pure supergravity were first obtained in
\cite{CSF78}. Moreover, equivalence with the $SO(4)$ theory was
shown, and the scalar and pseudoscalar of the theory were found to
parameterize the group $SU(1,1)$. This constitutes the first case
of appearance of hidden symmetries in supergravities. In
\cite{NT81} the pseudo-scalar was dualized to an antisymmetric
tensor, giving the theory vanishing trace anomaly. Thus a relation
with higher dimensional supergravities and string theory was
suspected.

This relation was further explored in \cite{Cha81a} where the
gravity multiplet of $N=1$ $d=10$ supergravity was dimensionally
reduced to obtain $N=4$ supergravity coupled to 6 Yang-Mills
vector multiplets. Upon reduction the metric and the tensor both
give 6 vectors $(g_{\mu})_i$ and $(B_{\mu})_i$ where $\mu$ is a
4-dimensional space-time index and $i$ is a 6-dimensional internal
index. The ``physical" vectors which are members of the multiplets
are linear combinations of these
\be
({\cal A}_{\mu})_i \  = \  (g_{\mu})_i  +  (B_{\mu})_i,\qquad
({\cal B}_{\mu})_i \  = \  (g_{\mu})_i  - (B_{\mu})_i
\ee
where
$({\cal A}_{\mu})_i$ are the graviphotons and $({\cal B}_{\mu})_i$
are the Yang-Mills vectors. The 36 scalars $g_{ij}$ and $B_{ij}$
go to a non-linear sigma model and the non compact symmetry is
extended to $SO(6,6)\times SU(1,1)$. One of the problems in this
situation is that the couplings between the graviphotons and the
Yang-Mills vectors are quite intricate, and a generalization to
any number n of Yang-Mills multiplets was not obvious. This was
later solved in \cite{dR85} for abelian multiplets, and in
\cite{dRo85}, \cite{dRW85}, \cite{Ter89} for non-abelian ones, using
conformal methods or more
specific reduction procedures.

However, a complete transcription of these results in superspace
formalism is still lacking. Our purpose in the present paper is to
attempt to (partially) fill in this gap. Among many advantages,
superspace provides concise and elegant descriptions of the
equivalent component theories, as well as a promising framework
for discovery of auxiliary fields in the hope of finding off-shell
formulations. $N=4$ pure supergravity multiplet with an
antisymmetric tensor (which we call N-T multiplet here) was first
obtained in \cite{NT81}. Its formulation in superspace encountered
a number of problems identified in \cite{Gat83} and overcome in
\cite{GD89}. Recently in \cite{GHK01}, the graviphotons were
identified in the central charge sector allowing an elegant
description of the Chern-Simons forms as coming from geometric
considerations. The equivalence with the component formalism was
proved in \cite{GK02} where the equations of motion coming from
the Bianchi Identities on superspace were shown to be identical to
the ones derived from the Lagrangian of \cite{NT81}. Considering
the historical developments in the component formalism, the next
``natural" step would be to look for $n$ abelian Yang-Mills
multiplets on the superspace. This is what we are presenting in
this work. We show that the central charge sector used in
\cite{GK02} can be extended to describe $n$ Yang-Mills vectors. In
section \ref{constraints} we recall the constraints allowing the
identification of the gravity multiplet and we give the
modifications we bring to the 0 dimensional scalar sector such
that the central charge sector can accommodate the extra matter
multiplets. $6n$ new scalars are found. In section
\ref{solutionBI}, we explore the implications of these new
constraints by giving the solution to the Bianchi Identities, and
we identify the $4n$ spin $1/2$ fermions belonging to the
Yang-Mills multiplets. In section \ref{identvect} we identify the
Yang-Mills vectors. In particular, we show that it is the geometry
of the 2-form, which, through the action of projectors, separates
the supergravity vectors from the Yang-Mills ones. Finally in
section \ref{discussion} we indicate how these results are
equivalent to the component formulation in \cite{dR85}. Then we
conclude in section \ref{outlook} by discussing a few
generalizations one can think of, such as describing non-abelian
Yang-Mills multiplets, or some hints for a possible off-shell
formulation.

\section{From pure supergravity to supergravity coupled to $n$
vector multiplets}

In this section we recall the essential results of \cite{GHK01}
concerning the identification of the components of the N-T
multiplet. Recall that in geometrical formulation of supergravity
theories the basic dynamic variables are chosen to be the vielbein
and the connection. Considering central charge superspace this
framework provides a unified geometric identification of graviton,
gravitini and graviphotons in the frame
$E^\ca=(E^a,E_\ta^\a,E^\ta_\da, E^\au)$
\be E^a\doubar\ =\ \vd x^m
e_m{}^a~,\quad E_\ta^\a\doubar\ =\ \frac{1}{2}\,\vd x^m
\p_m{}_\ta^\a~, \quad E^\ta_\da\doubar\ =\ \frac{1}{2}\,\vd x^m
\bar{\p}_m{}^\ta_\da~, \quad E^\au\doubar\ =\ \vd x^m v_m{}^\au~.
\eqn{frame}
Here $a$ is the vectorial index, $\a ,\da$ are the
spinorial indices, $\ta = 1..4$ is the internal symmetry index,
while $\au = 1..6+n$ counts the central charge coordinates.
Moreover, the
antisymmetric tensor can be identified in a superspace 2--form
$B$: \be B\doubar\ =\ \frac{1}{2}\,\vd x^m \vd x^n b_{nm}. \eqn{B}
The remaining component fields, a real scalar and 4 helicity 1/2
fields, are identified in the supersymmetry transforms of the
vielbein and 2--form, that is in torsion ($T^\ca=DE^\ca$) and
3--form ($H=dB$) components. The Bianchi identities satisfied by
these objects are
\be DT^\ca\ =\ E^\cb
R_\cb{}^\ca\,,\qquad\qquad\vd H\ =\ 0\,,
\ee
and, displaying the form coefficients
\bea
\left(_{\cd\cc\cb}{}^\ca\right)_T\quad&:&\quad E^\cb E^\cc E^\cd \left(\cd_\cd
T_{\cc\cb}{}^\ca+T_{\cd\cc}{}^\cf
T_{\cf\cb}{}^\ca-R_{\cd\cc\cb}{}^\ca\right)=0,
\label{ibT}\\
\left(_{\cd\cc\cb\ca}\right)_H\quad&:&\quad
E^\ca E^\cb E^\cc E^\cd \left(2\cd_\cd
H_{\cc\cb\ca}+3T_{\cd\cc}{}^\cf H_{\cf\cb\ca}\right)=0.
\label{ibH}
\eea

\subsection{The constraints}\label{constraints}

The geometrical description of the N-T multiplet is based on a set
of natural constraints in central charge superspace with structure
group $SL(2,\mathbb{C})\times U(4)$. The central charge sector is
chosen to be trivial in the sense that the covariant derivative in
the central charge direction $\cd_\au$ vanishes on all superfields
as well as the connection $\F_\az{}^\au$ is zero. Conventions for
vector and spinor representations of the Lorentz group are those
of \cite{BGG01}.

The generalizations of the
canonical dimension 0 ``trivial constraints" \cite{Mul86b} to
central charge superspace are
\be
           T{^\tc_\g}{^\tb_\b}{^a_{}}\ =\ 0~, \qquad
           T{^\tc_\g}{^\db_\tb}{^a}\ =\
-2i\d{^\tc_\tb}(\s{^a}\eps){_\g}{^\db}~, \qquad
           T{^\dg_\tc}{^\db_\tb}{^a}\ =\ 0~,
\eqn{T01} \be T{^\tc_\g}{^\tb_\b}{^\au}\ =\
4\eps_{\g\b}L^{1/2}\tt^{[\tc\tb]\au}~,\qquad
T^\tc_\g{}^\db_\tb{}^\au\ =\ 0~, \qquad
T{^\dg_\tc}{^\db_\tb}{^\au}\ =\
4\eps^{\dg\db}L^{1/2}\tt{_{[\tc\tb]}}{}^\au~.
\eqn{T02}

As explained in detail in the article \cite{GHK01}, the soldering
is achieved by requiring some analogous, ``mirror''-constraints
for the 2--form sector. Besides the -1/2 dimensional constraints
$H^\tc_\g{}^\tb_\b{}^\ta_\a = H^\tc_\g{}^\tb_\b{}^\da_\ta =
H^\tc_\g{}^\db_\tb{}^\da_\ta = H^\dg_\tc{}^\db_\tb{}^\da_\ta=0$,
we impose
\be
           H{^\tc_\g}{^\tb_\b}{_a}\ =\ 0~, \qquad
           H{^\tc_\g}{^\db_\tb}{_a}\ =\
-2i\d{^\tc_\tb}(\s{_a}\eps){_\g}{^\db}L~, \qquad
           H{^\dg_\tc}{^\db_\tb}{_a}\ =\ 0~,
\eqn{H01} \be
H{^\tb_\b}{^\ta_\a}_\au=4\eps_{\b\a}L^{1/2}\hh_\au{}^{[\tb\ta]}
\,,\qquad H^\tc_\g{}^\db_\tb{}_\au =0\,,\qquad
H^\db_\tb{}^\da_\ta{}_\au=4\eps^{\db\da}L^{1/2}\hh_{\au[\tb\ta]}\,,
\eqn{H02} with $L$ a real superfield. The physical scalar $\f$ of
the multiplet, called also graviscalar, is identified in this
superfield, parameterized as $L=e^{2\f}$. In turn, the helicity
1/2 gravigini fields are identified as usual \cite{How82},
\cite{GG83}, \cite{Gat83} in the 1/2--dimensional torsion
component \be \eps^{\b\g}T^\tc_\g{}^\tb_\b{}^\ta_\da\ =\
2T^{[\tc\tb\ta]}{}_\da,\qquad
\eps_{\db\dg}T_\tc^\dg{}_\tb^\db{}_\ta^\a\ =\
2T_{[\tc\tb\ta]}{}^\a. \ee

The scalar, the four helicity 1/2 fields, together with the gauge
fields defined in \equ{frame} and \equ{B} constitute the N-T
on-shell N=4 supergravity multiplet. Recall that in the case of
pure supergravity \cite{GHK01,GK02}, the matrix elements
$\tt^{[\tb\ta]\au},\tt_{[\tb\ta]}{}^{\au},\hh_{\au[\tb\ta]},
\hh_{\au}{}^{[\tb\ta]}$ are constrained to be covariantly constant
under the structure group $SL(2,\mathbb{C})\otimes SU(4)$.
However, by leaving them arbitrary, extra multiplets can be
accommodated in the same geometry \cite{Gat83}. In particular,
imposing the self--duality conditions
\be \tt^{[\td\tc]\au} \ =\
\frac{q}{2}\,
\e^{\td\tc\tb\ta}\tt_{[\tb\ta]}{}^\au\,,\qquad\textrm{with}\quad
q=\pm1
\eqn{du0}
we expect to describe a number of on-shell vector
multiplets \cite{Soh78a} coupled to the N-T multiplet. Recall,
   that the central charge index, $\au$, runs from 1 to $6+n$. Since
the number of
gauge vectors taking part of the N-T multiplet is $6$, we expect to
deal implicitly with $n$ independent additional gauge vectors in the
geometry, which take part of vector multiplets.

Like in the case of pure N-T supergravity, let us suppose the
existence of a covariantly constant metric $\gg^{\az\au}$ \be
\gg^{\az\au}\gg_{\au\av}=\d^\az_\av, \ee which connects the
components of the 2-form $T^\au$ to the components of $H$ having
at least one central charge index
\bea
H_{\az\cd\cc}&=&T_{\cd\cc}{}^\au\gg_{\au\az}.
\label{relTH}
\eea
However, unlike the pure supergravity case, this metric is not entirely
given as a function of the Lorentz scalars $\tt^{[\tb\ta]\au}$ or
$\hh_\au{}^{[\tb\ta]}$, which are in this case $6\times(6+n)$
matrices having at most
rank $6$.

One can further eliminate a big number of superfluous fields  by
assuming the constraint \be T_{\az\cb}{}^\ca=0, \eqn{tz} as well
as all possible compatible conventional constraints. Also, the
constraints
\be
\cd^\te_\a(T^{[\td\tc]\au})H_{\au[\tb\ta]}\ =\ 0\,,\qquad\qquad
\cd_\te^\da(T_{[\td\tc]\au})H^{\au[\tb\ta]}\ =\ 0
\ee
at dimension 1/2 as well as
\be
\bar\s^{b\da\g}T_{\g}^\tc{}_b{}_\da^\ta = 0
\,,\qquad\qquad \s_{\a\dg}^bT^{\dg}_\tc{}_b{}^\a_\ta = 0\,,
\eqn{dt}
at dimension 1 are used to put the gravity part on--shell.

Finally, in order to make possible the direct comparison with the results of
pure N-T supergravity, let us define a covariant derivative $\hat\cd$ under
$SL(2,\mathbb{C})\times SU(4)$,
\bea
\cd v^\ta&=&\hat{\cd}v^\ta-\chi^\ta{}_\tb v^\tb ,
\eea
where the shift $\chi^\ta{}_\tb$ in the connection is determined by
the requirement
\be
\hat{\cd} (\tt^{[\td\tc]\au})\hh_{\au[\tb\ta]}\ =\ 0.
\eqn{Dth}

\subsection{Solution of the Bianchi identities}\label{solutionBI}

Now let us consider the torsion and 3--form $H$ subject to the
above summarized constraints, and look at the Bianchi Identities
\equ{ibT} and \equ{ibH} as equations for the remaining components
of these two objects. The solution of these Bianchi Identities
will be presented in the order of growing canonical dimension.

The lowest canonical dimension Bianchi Identities are those with
only spinorial indices written for the 3--form. Given the above
constraints, they are satisfied if and only if the Lorentz scalars
at dimension 0 satisfy
\be
\gg_{\au\az}\tt^{[\td\tc]\au}\tt_{[\tb\ta]}{}^\az\ =\
\frac{1}{2}\d^{\td\tc}_{\tb\ta}.
\eqn{th}
One may recognize that
they represent $6\times6$ equations for the $6\times(6+n)$ a priori independent
scalar superfields $\tt^{[\td\tc]\au}$. This means, that there are
$6\times n$ degrees of freedom left in these scalars, which is exactly the
number of scalars in $n$ additional vector multiplets.

At dimension 1/2 the spinorial components of $\chi$
are obtained and they are found to be the same as in the pure N-T
case \be \ki_\a^\ta{}^\tb{}_\tc\ =\ \frac{1}{4}\d^\tb_\tc
\ki_\a^\ta,\qquad \ki^\da_\ta{}^\tb{}_\tc\ =\
-\frac{1}{4}\d^\tb_\tc \ki^\da_\ta, \ee where we used the notation
\mbox{$\ki_\a^\ta=L^{-1}\cd_\a^\ta L$},
$\ki^\da_\ta=L^{-1}\cd^\da_\ta L$.

Then the solution of the Bianchi Identities at dimension 1/2 can be written
in the following way
\bea
T_{[\tc\tb\ta]\a}&=&q\e_{\tc\tb\ta\tf}\ki^\tf_\a \\
\hat{\cd}^\td_\d
\tt^{[\tc\tb]\au}&=&q\e^{\td\tc\tb\ta}\xx_{\d\ta}{}^\au\label{d12a}\\
T^\dg_\tc{}_b{}^\au&=&\vi (\bs_b)^{\dg\a}\left(\xx_{\a\ta}{}^\au+\ki^\tf_\a
\tt_{[\tf\ta]}{}^\au\right)L^{1/2}\label{d12b}
\eea
while similar relations are implied for $H$
which can be easily obtained using \equ{relTH}
\bea
\hat{\cd}^\td_\d
\hh_\au{}^{[\tc\tb]}&=&q\e^{\td\tc\tb\ta}\gg_{\au\az}\xx_{\d\ta}{}^\az\\
H^\dg_\tc{}_b{}_\au&=&\vi
(\bs_b)^{\dg\a}\gg_{\au\az}\left(\xx_{\a\ta}{}^\az+\ki^\tf_\a
\tt_{[\tf\ta]}{}^\az\right)L^{1/2}. \eea Inspecting these results
one identifies the spinor of the supergravity multiplet in
$T_{[\tc\tb\ta]\a}$ as the lowest superfield component of
$\ki^\ta_\a$, while the spinor $\xx_{\a\ta}{}^\au$ appearing in
the spinorial derivative of the scalars has to take part of the
$n$ additional vector multiplets. Their number is
$4*(6+n)-4*6=4*n$, where the number of 4*6 independent relations
in the spinorial component of \equ{Dth} is considered.

The results at dimension 1 can be gathered in three sectors
depending on the irreducible representation of the Lorentz group
in which the double spinorial derivatives on the 0 dimensional
scalar sits.

Let us
start with the sector of the scalars and of the dual fieldstrength
$H_a^*=\frac{1}{3!}\e_{abcd}H^{bcd}$ of the antisymmetric tensor.
This sector is determined by the mixed derivatives on the 0
dimensional scalars: \be K_\b{}^\da{}^\tb{}_\ta\ =\
[\cd^\tb_\b,\cd^\da_\ta]L,\qquad \cd^\td_\d \xx^{\tc\db \au}. \ee
The Bianchi Identity $\br{_\d^\td{}^\dg_\tc{}_{ba}}$ is satisfied
if and only if
\be \frac{1}{2} K_a{}^\td{}_\tc +\d^\td_\tc H^*_a +4\vi
U_a{}^\td{}_\tc L \ =\ \gg_{\az\au}\xx_{\tc}{}^\az\s_a
\xx^{\td\au}L +\frac{1}{2}\d^{\td\te}_{\tc\tf}\ki^\tf\s_a\ki_\te
L. \ee This equation relates the commutator on $L$ to the dual
fieldstrength $H_a^*$ of the antisymmetric tensor and the
superfield $U_a{}^\td{}_\tc$, which appears in the dimension 1
torsion components \bea
T^\tc_\g{}_b{}^\a_\ta&=&-2(\s_{ba})_\g{}^\a U^a{}^\tc{}_\ta\\[2mm]
T_\tc^\dg{}_b{}_\da^\ta&=&2(\bs_{ba})^\dg{}_\da U^a{}^\ta{}_\tc.
\eea

Also, the Bianchi Identity
$\brg{_\d^\td{}^\dg_\tc{}_\b^\tb{}^\a_\ta}$ together with the
definition of the $SU(4)$ covariant derivative gives an other
independent relation for the commutator on $L$ \be
K_a{}^\td{}_\tc\ =\ 2\vi \d^\td_\tc(U_a+\ki_a)L
-(\d^\td_\tb{}^\ta_\tc+\d^\td_\tb\d^\ta_\tc) \ki^\tb\s_a \ki_\ta
L. \ee

The last Bianchi Identities relevant for this sector are carrying a central
charge index, $\brg{_\d^\td{}^\tc_\g{}_b{}^\au}$ and
$\br{_\d^\td{}^\tc_\g{}_b{}^\au}$. Using \equ{Dth}, these imply the equations
\be
K_a\ =\ 8\vi \ki_a L +\ki^\ta\s_a\ki_\ta L
-2\gg_{\az\au}\xx_\ta{}^\az\s_a\xx^\ta{}^\au L
\eqn{tr3}
\be
\widetilde{\ki}_a{}^\tb{}_\ta\ =\ -\frac{\vi}{8}
\left(\widetilde{\ki^\tb\s_a\ki_\ta}
+2\gg_{\az\au}\widetilde{\xx_\ta{}^\az\s_a\xx^\tb{}^\au}\right)
\eqn{str3}
where no indices are written for the trace parts and the tilde denotes the
traceless parts.

It is sufficient then to solve the system of equations for the trace part
\bea
U_a&=&\frac{\vi}{8}\left(\ki^\ta\s_a\ki_\ta
+2\gg_{\az\au}\xx_\ta{}^\az\s_a\xx^\ta{}^\au\right)\\
K_a&=&-8H_a^*
+4\ki^\ta\s_a\ki_\ta L
+4\gg_{\az\au}\xx_\ta{}^\az\s_a\xx^\ta{}^\au L\\
\ki_a&=&\vi H_a^*L^{-1}
-\frac{3\vi}{8}\left(\ki^\ta\s_a\ki_\ta
+2\gg_{\az\au}\xx_\ta{}^\az\s_a\xx^\ta{}^\au\right)
\eea
and for the traceless part
\bea
\widetilde{U}_a{}^\tb{}_\ta& =& \widetilde{\ki}_a{}^\tb{}_\ta\ =\
-\frac{\vi}{8}
\left(\widetilde{\ki^\tb\s_a\ki_\ta}
+2\gg_{\az\au}\widetilde{\xx_\ta{}^\az\s_a\xx^\tb{}^\au}\right)\\
\widetilde{K}_a{}^\tb{}_\ta&=& -2\widetilde{\ki^\tb\s_a\ki_\ta} L
\eea
in order to have all the objects of this sector
expressed in terms of the dual fieldstrengths
$H_a^*$ and nonlinear terms in the spinors $\ki$ of the gravity and
$\xx$ of the
vector multiplets.

Notice, that the vector $P_a$ which appears naturally in the
superspace geometry
\be
\cd_\d^\td T^{[\tc\tb\ta]\da}\ =\ q\e^{\td\tc\tb\ta}P_\d{}^\da
\eqn{P}
is also corrected by non-linear terms in the $\xx$ spinors of the additional
vector multiplets
\bea
P_a& =& \vi L^{-1}\cd_aL
+L^{-1}H_a^*
-\frac{3}{4}\ki^\ta\s_a\ki_\ta
-\frac{1}{2}\gg_{\az\au}\xx_\ta{}^\az\s_a\xx^\ta{}^\au
\eea

Finally, using all these results, one obtains for the mixed
derivatives of $\xx$
\be
\hat{\cd}^\td_\d \xx^{\tc\db\au}\ =\
-2\vi\hat{\cd}_\d{}^\db\tt^{[\td\tc]\au}
+2\gg_{\az\av}\xx_{\d\tf}{}^\az \xx^{\tc\db\av} \tt^{[\tf\td]\au}
-\frac{1}{4}\ki^\td_\d\xx^{\tc\db\au}.
\ee

The second sector at dimension 1 is the sector of the gauge vectors
$v_m{}^\au$,
with fieldstrengths
identified in $F_{ba}{}^\au\doteq T_{ba}{}^\au$. This sector is
governed by the double derivatives which are symmetric in their
spinorial indices
\be
\cd^\tb_{(\b}\cd^\ta_{\a)}L,\qquad
\cd^\td_{(\d} \xx_{\tc\b)}{}^{\au}.
\ee
After an analysis of the relevant Bianchi Identities one finds
\bea
\cd^\tb_{(\b}\cd^\ta_{\a)}L&=&-4F_{(\b\a)}{}^\au\hh_\au{}^{[\tb\ta]}L^{1/2}
+qL\epsilon^{\tb\ta\te\tf}\gg_{\au\az}\xx_{(\b\te}^\au
\xx_{\a)\tf}^\az
\label{DDLsym}\\
G_{(\b\a)[\tb\ta]}&=&-2\vi F_{(\b\a)}{}^\au\hh_\au{}_{[\tb\ta]}L^{-1/2}
+\vi\gg_{\au\az}\xx_{(\b\tb}^\au \xx_{\a)\ta}^\az \label{Gdown}
\\
G_{(\b\a)}{}^{[\tb\ta]}&=&
-\vi\frac{q}{2}
\e^{\tb\ta\te\tf}\gg_{\au\az}\xx_{(\b\te}^\au \xx_{\a)\tf}^\az \label{Gup}
\eea
where $G_{(\b\a)[\tb\ta]}$ and $G_{(\b\a)}{}^{[\tb\ta]}$ are the selfdual parts
of the antisymmetric tensors $G_{ba[\tb\ta]}$ and $G_{ba}{}^{[\tb\ta]}$
appearing in the torsion components
\bea
T_\tc^\dg{}_b{}_\ta^\a&=&\frac{1}{2}(\bs^f)^{\dg\a}G_{bf}{}_{[\tc\ta]}\label{TF}\\[2mm]
T^\tc_\g{}_b{}^\ta_\da&=&\frac{1}{2}(\s^f)_{\g\da}G_{bf}{}^{[\tc\ta]}.
\eea
Then the derivative of the spinor $\xx$ becomes
\bea
\hat{\cd}^\td_{(\d}\xx_{\b)\tc}{}^\au&=&
2\d^\td_\tc\left(F_{(\d\b)}{}^\au
-F_{(\d\b)}{}^\az\hh_{\az[\tb\ta]}\tt^{[\tb\ta]\au}\right)L^{-1/2}\nn\\
&&+2\gg_{\av\az}\xx^\av_{(\d\tb}\xx_{\b)\tc}^\az\tt^{[\tb\td]\au}
+\frac{1}{4}\ki^\td_{(\d}\xx_{\b)\tc}^\au
-\d^\td_\tc\ki^\tf_{(\d}\xx_{\b)\tf}^\au\,.
\label{Dx}
\eea

The last sector might be called the sector of auxiliary fields, which in the
present on-shell case does not contain any new superfield. It is governed by
the antisymmetric part of the double derivatives
\be
\cd^{\ta\a}\cd_\a^\tb L,\qquad\cd^{\ta\a}\xx_{\a\tb}{}^\au.
\ee
Here one finds the relations
\bea
\cd^{\ta\a}\cd_\a^\tb L&=&-2\gg_{\az\au}\xx^{\ta\az}\xx^{\tb\au}L\label{DDL}\\
\hat{\cd}^{\ta\a}\xx_{\a\tb}{}^\au&=&
        2\gg_{\av\az}\xx_\tf{}^\av\xx_\tb{}^\az \tt^{[\tf\ta]\au}
+\frac{1}{4}\ki^\ta\xx_\tb{}^\au
-2\ki_\tb\xx^{\ta\au}
+\d^\ta_\tb\ki_\tf\xx^{\tf\au}
\eea
as well as that the central charge component of the shift in the connection,
$\ki_\au{}^\tb{}_\ta$, has to vanish.

\subsection{Identification of the component
fields}\label{identvect}

To sum up, let us review the main results obtained so far. The
identification of the N-T supergravity multiplet in the geometry
goes exactly in the same manner as in the  pure supergravity case
\cite{GHK01, GK02}.  The difference arises from the fact that
whereas the Lorentz scalars $\tt^{[\tb\ta]\au}$ sitting in
the 0 dimensional torsion components in the central charge
direction are covariantly constant for the pure supergravity case,
they are taken to be a priori general superfields here. It turns
out that they belong to the only matter multiplets available in
$N=4$ supergravity, that are vector multiplets. Considering that
they obey the $6\times6$ independent relations (\ref{th}), we obtain
$6\times n$
scalars in $n$ vector multiplets. Remark that (\ref{th}) is
precisely the relation obtained in \cite{dR85} as a condition for
breaking conformal symmetry and going to Poincar\'e gauge.
However, notice, that in our approach the equation \equ{th}
satisfied by the scalars is a consequence of the Bianchi
Identities of a 3--form H constrained in an analogous way as we
did for the pure N-T supergravity. In particular, the constraint
\equ{relTH}, which relates components of H to those of the torsion
by a metric plays a key role in all the basic features of the
coupled system.

As usual in the superspace formalism, supersymmetry generators are
implemented as spinorial covariant derivatives. Therefore we
expect to see the fermionic supersymmetric partners of the
scalars, $\xx_{\a\ta}{}^\au$, arising in the object
$\hat\cd_{\a}^\td\tt^{[\tc\tb]\au}$. This is indeed the case as
shown by equation (\ref{d12a}), whereas the constraint (\ref{Dth})
insures that only the right number, $n\times 4$, of them are
non-vanishing. Further applying spinorial derivatives to the
spinors $\xx_{\a\ta}{}^\au$, we should see the fieldstrength of
Yang-Mills vectors. Accordingly, \equ{Dx} suggests to define \be
F^\ym_{ba}{}^\au\ =\
F_{ba}{}^\az\left(\d^\au_\az-\hh_{\az[\tb\ta]}\tt^{[\tb\ta]\au}\right).
\eqn{fym}
       Also, let us identify the gauge vectors of the supergravity
multiplet, the graviphotons, as those which appear in the
supersymmetry transformation of the spinor $\ki$ which takes part
of the supergravity multiplet \equ{DDLsym}:
\be
F^\sg_{ba}{}^\au\
=\ F_{ba}{}^\az\hh_{\az[\tb\ta]}\tt^{[\tb\ta]\au}.
\eqn{fsg}
One might
notice that the two definitions, \equ{fym} and \equ{fsg}, involve projectors
on the fields
which belong to the supergravity or the gauge multiplets: \be
P^\sg{}_\az{}^\au\ =\ \hh_{\az[\tb\ta]}\tt^{[\tb\ta]\au},\qquad
P^\ym{}_\az{}^\au\ =\ \d^\au_\az-P^\sg{}_\az{}^\au. \ee Due to the
identity \equ{th} they possess the standard properties of
projectors
\be
(P^\sg)^2\ =\ P^\sg,\qquad(P^\ym)^2\ =\ P^\ym,\qquad P^\sg
P^\ym\ =\ 0.
\ee
The dimension of the spaces on which they project
can also be computed and found to be as expected
\be
\textrm{tr}P^\sg\ =\ 6\,\qquad\quad\textrm{tr}P^\ym\ =\ n.
\ee

Let us then
summarize the identification of the fields in the following table:
\begin{center}
\begin{tabular}{||c|l|l||}
\hline
         &  & \\
$\s$ & N-T sugra multiplet        & $n$ gauge multiplets  \\
         &  & \\
         \hline
                  &  & \\
2  &$e_m{}^a$ &\\
         &  & \\
3/2&$\p_m{}^\a_\ta$&\\
         &  & \\
1&$v_m^\sg{}^\au,\quad F^\sg_{ba}{}^\au P^\ym_\au{}^\az\loco=0$&
$v_m^\ym{}^\au,\quad F^\ym_{ba}{}^\au P^\sg_\au{}^\az\loco=0$\\
         &  & \\
1/2&$\ki^\ta_\a\loco$&
$\xx_{\a\ta}{}^\au\loco,\quad\xx_{\a\ta}{}^\au P^\sg_\au{}^\az\loco=0$\\
         &  & \\
0$_s$+0$_t$& $L\loco$ and $b_{mn}$ & $\tt^{[\tb\ta]\au}\loco,\quad
\gg_{\au\az}\tt^{[\td\tc]\au}\tt_{[\tb\ta]}{}^\az\loco=\frac{1}{2}\d^{\td\tc}_{\tb\ta}$\\
         &  &\\
         \hline
\end{tabular}
\end{center}

\section{Discussion}\label{discussion}

To give further arguments for the equivalence of this formulation
with the component formalism, we discuss now the modifications
found in the equations of motion due to the emergence of the
Yang-Mills sector. One of the most intriguing features of the
above results is the correction of the antisymmetric part of the
double derivative on L \equ{DDL} by the quadratic term in the
gaugini. This term appears in the corresponding derivative of the
gravigino superfield \bea
\cd^{\td\a}T_{[\tc\tb\ta]\a}&=&q\e_{\tc\tb\ta\tf}L^{-1}\cd^{\b\td}\cd_\b^\tf
L\nn\\
&=&-2q\e_{\tc\tb\ta\tf}\gg_{\az\au}\xx^{\td\az}\xx^{\tf\au}L. \eea
It is well known that the object $\cd^{\td\a}T_{[\tc\tb\ta]\a}$ is
playing the role
of an auxiliary field%
\footnote{The lowest component of this auxiliary superfield
is the $E^{ij}$ auxiliary field \cite{BdRdW81} in the component
description of off-shell
conformal $N=4$ supergravity.} in four dimensional $N=4$ conformal supergravity
\cite{GG83}, \cite{CL86}.
It was explicitly shown in \cite{GK02} how its
vanishing implies the Dirac equation for the gravigino and at dimension 2
the equations of motion for the antisymmetric tensor and that of the scalar.

Let us do the exercise of deriving equations of motion in an
analogous way as in the pure supergravity case \cite{GK02}.
However, this time we do not consider all the non-linear terms,
but concentrate only on the main features of the coupling keeping
only terms involving the fieldstrength of gauge vectors. In
particular, using \equ{P} one can write the identity \be
\sum_{\td\tc}\left(\left\{\cd^\te_\e,\cd_\td^\dd\right\}
T_{[\tc\tb\ta]\a} -\cd_\td^\dd\left(\cd^\te_\e
T_{[\tc\tb\ta]\a}\right)\right)\ =\ 0. \ee Observe that the
antisymmetric part of this relation in the indices $\e$ and $\a$
gives rise to Dirac equation for the helicity 1/2 fields, that is
$\partial^{\a\dd}T_{[\tc\tb\ta]\a}=0$ in the linear approach.
However, this time $\cd^{\te\a} T_{[\tc\tb\ta]\a}$ is different
from zero and gives rise to a fieldstrength of the Yang-Mills
vectors by \equ{Dx} when the spinor derivative acts on the spinors
$\xx$ \be
\partial^{\a\dd}\ki^\ta_\a\ =\
2\vi
F^{(\dd\da)}{}^\av\xx_\da^{\ta\au}P^\ym{}_\av{}^\az\gg_{\az\au}L^{-1/2}+...
\ee
After further differentiating by $\cd_{\dd\ta}$, in order to
obtain the expression of $\Box L$ one will need to use the algebra of
derivatives on superspace. This operation (intimately connected to
the gravity part) involves torsion components of \equ{TF} near the
spinorial derivative on $\ki^\ta_\a$ \equ{DDLsym}, both containing
the vectors of the supergravity multiplet:
\be
\Box L\ =\
-\vi\partial^aH_a^* +2\left(F^{(\da\db)\au} F_{(\da\db)}{}^\az
P^\ym{}_\au{}^\av -F^{(\a\b)\au} F_{(\a\b)}{}^\az
P^\sg{}_\au{}^\av\right)\gg_{\av\az}+...
\eqn{bL}

On the one hand, taking the real part of this relation one finds
the equation of motion for the scalar \be \Box L\ =\
-\frac{1}{2}F^{ba\au} F_{ba}{}^\az
\left(P^\sg-P^\ym\right)_\au{}^\av\gg_{\av\az}+...
\eqn{eomL}
and may conclude that indeed, both the kinetic term for the
graviphotons and the Yang-Mills vectors have to be present in the
Lagrangian of the theory. Moreover, the coupling matrix of the
kinetic term for the vectors
\be
\OM_{\au\az}\ =\ \left(P^\sg-P^\ym\right)_\au{}^\av\gg_{\av\az}
\ee is
       exactly the one appearing in the action exhibited in \cite{dR85}.
       There it was argued that, in order for all vectors to be
       physical, the above coupling must be positive definite, which
       implies that the metric $\gg_{\au\az}$ must have signature $(6,n)$.
This additional information about the metric together with the
equation \equ{th} satisfied by the scalars allows to identify the
sigma model $\frac{SO(6,n)}{SO(6)\times SO(n)}$ parameterized by
the Yang-Mills scalars $\tt^{[\tb\ta]}{}^{\az}$.  The
identification of this sigma model first suggested in
\cite{Cha81a} and discussed also in \cite{dR85} is detailed in
appendix \ref{sigma}.

On the other hand the imaginary part of the relation \equ{bL}
gives the Bianchi Identity for the 2-form gauge field \be
\partial^aH_a^*\ =\ \frac{\vi}{2}F^*{}^{ba\au} F_{ba}{}^\az
\left(P^\ym+P^\sg\right)_\au{}^\av\gg_{\av\az}+... \ =\
\frac{\vi}{2}F^*{}^{ba\au} F_{ba}{}^\az\gg_{\au\az}+... \eqn{biH}
The topological term $F^*{}^{ba\au} F_{ba}{}^\az\gg_{\au\az}$ is
an indication of the intrinsic presence of Chern-Simons forms in
the theory. Indeed, one needs just to explicit the projection
$E^\ca\doubar = e^\ca = \vd x^m e_m{}^\ca$ on the 3-form $H$ \bea
H\doubar& =& \frac{1}{2}\vd x^m\vd x^n\vd x^k \partial_kb_{nm}\ =\
\frac{1}{3!}e^\ca e^\cb e^\cc H_{\cc\cb\ca}\loco\nn\\
&=&\frac{1}{3!}e^a e^b e^c H_{cba}+\frac{1}{2}e^a e^b
\vd x^m v_m{}^\au H_{\au ba}+...\nn\\
&=&\frac{1}{3!}e^a e^b e^c H_{cba}+\frac{1}{2}e^a e^b \vd x^m
v_m{}^\au F_{ba}{}^\az\gg_{\au\az}+...
\eea
where for the last
line the relation
\equ{relTH} was used, in order to see that in the
development of the supercovariant fieldstrength of the
antisymmetric tensor $H^*_{a}\loco$ the fieldstrength
$e_l{}^a\e^{knml}\partial_{k}b_{nm}$ is naturally accompanied
by the Chern-Simons terms of both the graviphotons and the
additional gauge vectors. This is an intrinsic property of
soldering in central charge superspace, as pointed out in
\cite{GHK01}. One can clearly see that it is
the existence of an object $\gg$ in the central charge sector
relating the 3--form components to those of the torsion by \equ{relTH},
which is responsible for this issue.

Recall, that in constructions of coupling of supergravity
containing an antisymmetric tensor to Yang-Mills multiplets
\cite{BGG01}, the usual procedure is to define a modified
fieldstrength for the antisymmetric tensor including by hand the
Yang-Mills Chern-Simons terms in it. In this case, the gauge
transformations of the Chern-Simons term are compensated by
assigning suitably adjusted Yang-Mills gauge transformations to
the antisymmetric tensor, and the modified fieldstrength is
rendered invariant in this way. Let us verify, that this is
automatical in our approach using central charge superspace. Here
gauge transformations are identified as translations in the
direction of the central charge coordinates. Indeed, taking the
double bar projection \cite{BGG01} for the Wess-Zumino
transformation of the frame component $E^\au$ along the vector
field $\z^\ca=(0,0,0,\z^\au)$ \be \d^\wz_\z E^\au\ =\
D\z^\au+\imath_\z T^\au, \eqn{vtf} one finds the usual
transformation law for abelian gauge vectors \be \d^\wz_\z
v_m{}^\au\ =\ \partial_m\z^\au. \ee However, writing the
Wess-Zumino transformation for the 2--form gauge potential \be
\d^\wz_\z B\ =\ \imath_\z H\ =\ \frac{1}{2}E^\ca E^\cb \z^\au
H_{\au\cb\ca} \ = \ \frac{1}{2}E^\ca E^\cb \z^\au
\gg_{\au\az}T_{\cb\ca}{}^\az \ =\ \gg_{\au\az}\z^\au T^\az \ee and
taking its double bar projection one finds that the antisymmetric
tensor transforms exactly into the fieldstrengths of the gauge
vectors \be \d^\wz_\z b_{mn}\ =\ \gg_{\au\az}\z^\au
\left(\partial_{m}v_{n}{}^\az-\partial_{n}v_{m}{}^\az\right). \ee
where $T^\az\doubar=\frac{1}{2}\vd x^m\vd
x^n(\partial_{n}v_{m}{}^\az -\partial_{m}v_{n}{}^\az)$ was used.

\section{Conclusion and outlook}\label{outlook}

In this article we identified the N-T multiplet coupled to $n$
abelian vector multiplets in the geometry of central charge
superspace. Even though we started with $6+n$ gauge vectors, the
geometry of the 3--form singled out the particular combinations of
these which belong to the supergravity multiplet. The remaining
independent combinations take part of the additional gauge
multiplets. The supersymmetry transformations as well as main
parts of some equations of motion are compared to the component
formulations found by dimensional reduction \cite{Cha81a} or using
conformal methods \cite{dR85}. In particular, we saw the emergence
of the $SO(6,n)/SO(6)\times SO(n)$ sigma model for the Yang-Mills
scalars as well as the presence of Chern-Simons terms in the
supercovariant fieldstrength of the antisymmetric tensor or its
particular transformation under Yang-Mills gauge transformations.
On a more technical level we also could point out that the
quadratic term in the gaugini sitting at the place of an auxiliary
field of conformal supergravity was crucial.

Let us emphasize here that the features of the coupling pointed
out in this article are very general properties of coupling
supergravity containing an antisymmetric tensor with Yang-Mills
gauge theory \cite{CM83}, currently used as guiding principles in
superspace descriptions. In the articles \cite{BGG01} and
\cite{GG99} one can find a review of this kind of couplings in the
four dimensional $N=1$ case. Also, an extensive list of references
concerning various constructions of coupled systems of the same
type in higher dimensions can be found in the same articles. Let
us take for example \cite{BDG90}. There the aim was to incorporate
string corrections up to first order in the string slope-parameter
in the ten dimensional $N=1$ superspace. It turned out that the
inclusion of both the Yang-Mills and Lorentz Chern-Simons terms in
the geometry goes hand-in-hand with the presence of a source \be
A_{cba}\ \sim\ \b^\prime\textrm{tr}(\la\s_{cba}\la)
+\g^\prime(T_{kl}\s_{cba}T^{kl}), \ee which appears in the double
spinorial derivative of the dilaton and plays the role of an
auxiliary field in the pure ten dimensional $N=1$ supergravity.
Observe, that $A_{cba}$ is quadratic in the gaugino fields and the
gravitino fieldstrength, and one can show in a similar way as we
did above, that it is responsible for the curvature squared terms
in the corresponding Lagrangian.

However, the interest of the central charge superspace approach
applied here to the $N=4$ case in four dimensions is that all the
features of the Chapline--Manton coupling come out automatically offering us
the possibility to just study the underlying mechanisms.

Considering possible generalisations of the work presented here,
an obvious next step would be to check whether non-abelian vector
multiplets can be described in this framework. Let us go back to
the transformation law \equ{vtf} of the frame component $E^\au$ in
which the gauge vectors are identified and take its double
projection in a more general setup 
\be 
\d^\wz_\z v_m{}^\au\ =\
\partial_m\z^\au\ +\ v_m{}^\av\z^\az T_{\az\av}{}^\au\ +\ ...
\eqn{v} 
Then the torsion component superfield $T_{\az\av}{}^\au$
-- vanishing in the present work -- can play the role of structure
constants and one can interpret the above equation as the
transformation law for non-abelian gauge vectors. It would be
interesting in particular to investigate whether superspace
geometry implies some restrictions on the possible Lie groups.

\acknowledgments{We would like to thank S.J. Gates, Jr. and R. Grimm for
their advice, as well as for their comments and suggestions after
reading the manuscript. The work of S.G. was supported by the Japanese
Society for the Promotion of Science (JSPS) under contract
P-03743. Research of A.K. was supported by the Center for String and
Particle Theory, UMD and by the National Science Foundation 
grant PHY-01523911.}

\begin{appendix}
\section{Supersymmetry transformations}
\label{tfs}

Let us sum up in this appendix the supersymmetry transformations
of the component fields. In superspace description of supergravity
theories these are encoded in the formulas of Wess-Zumino
transformations along a vector field $\x$
  \bea
\d^\wz_\x E^\ca&=&\imath_\x T+D\x^\ca\\
\d^\wz_\x B&=&\imath_\x H\\
\d^\wz_\x \om&=&\imath_\x D\om \eea where $\om$ is a covariant
superfield. Considering $\x^\ca=(0,\x^\a_\ta,0,0)$ one finds for
the component fields of the supergravity multiplet \bea \d^\wz_\x
e_m{}^a & = & \vi \x_\tc\s^a \bar\psi_m{}^\tc\\
\frac{1}{2}\d^\wz_\x \p_m{}^\a_\ta & = & \hat{\cd}_m \x^\a_\ta
-2(\x_\tc\s_{ma})^\a U^{a\tc}{}_\ta\loco
+\x^\a_\tc e_m{}^a \ki_a{}^\tc{}_\ta\loco\nn\\
&&+\frac{1}{8}\x^\a_\ta\left(\p_{m\tb}\ki^\tb-\bp_m{}^\tb\ki_\tb\right)
-\frac{1}{8}(\x_\tb\ki^\tb)\p_m{}^\a_\ta\\
\frac{1}{2}\d^\wz_\x \bp_m{}_\da^\ta & = & \vi e_m{}^b
(\x_\tc\s^a)_{\da}
F^{\sg(-)}_{ab}{}^\az\tt^{[\tc\ta]\au}\gg_{\az\au}L^{-1/2}\nn\\
&& +\frac{q}{2}\psi_{m\tf}\x_\tc\epsilon^{\tf\tc\ta\tb}\bar\chi_{\tb\da}
+\frac{1}{8}\bp_m{}^\ta_\da(\x_\tc\ki^\tc)\nn\\
&&-\frac{\vi}{2}\left((\x_\tb\s_m\xx^{[\ta\az})\xx_\da^{\tb]\au}
+\frac{q}{2}\e^{\ta\tb\te\tf}(\x_\tb\xx_\te{}^\az)(\xx_\tf{}^\au\s_m)_\da\right)
\gg_{\au\az}\\
\left(\d^\wz_\x v_m{}^\au\right)P^\sg{}_\au{}^\az & = &
2L^{1/2}\psi_{m\tb}\x_\tc \tt^{[\tb\tc]\az}
-\vi L^{1/2}(\x_\tc\s_m\bar\chi_\tb )\tt^{[\tc\tb]\az}\\
\d^\wz_\x b_{mn} & = & 2\vi L^{1/2} g_{\au\az}v_{[n}{}^\au
(\x_\tc\s_{m]})_{\da}\left(\xx^{\da\tc}{}^\az+\ki_\tf^\da\tt^{[\tf\tc]\az}\right)\
\nn\\
&&+2L\x_\tc\s_{mn}\chi^\tc
+2\vi L\bar\psi_{[n}^\tc\s_{m]}\x_\tc
+4L^{1/2}v_{[m}{}^\au\psi_{n]\tf}\x_\tc\hh_\au{}^{[\tf\tc]}\\
\d^\wz_\x L & = & \x_\tc \ki^\tc L\\
\d^\wz_\x \ki_\a^\ta & = & -2(\s^{ab}\x_\tc)_\a F^\sg_{ab}{}^\az\tt^{[\tc\ta]\au}\gg_{\az\au}\nn\\
&& +\left(\x_{\a\tc}(\xx^{\tc\az}\xx^{\ta\au})
+q\e^{\tc\ta\te\tf}(\x_\tc\xx_\te{}^\au)\xx_{\a\tf}{}^\az\right)\gg_{\az\au}
-\frac{3}{4}(\x_\tc\ki^\tc)\ki^\ta_\a\\
\d^\wz_\x \ki^\da_\ta & = &
\vi(\x_\ta\s^a\eps)^\da\left(L^{-1}\cd_a L-U_a-\ki_a\right)\loco
+\frac{3}{4}(\x_\tb\ki^\tb)\ki^\da_\ta-(\x_\ta\ki^\tb)\ki^\da_\tb
\eea 
For the components of the vector multiplets one obtains 
\bea
\d^\wz_\x \tt^{[\tb\ta]\au}&=&
q\e^{\tb\ta\td\tc}\x_\td\xx_\tc{}^\au\\
\d^\wz_\x \xx_{\a\ta}{}^\au&=&
(\s^{ab}\x_\ta)_\a F^\ym_{ab}{}^\au  L^{-1/2}
-2(\x_\tc\xx_\tb{}^\av)\xx_{\a\ta}{}^\az \tt^{[\tc\tb]\au}\gg_{\av\az}
+\frac{1}{4}(\x_\tb\ki^\tb)\xx_{\a\ta}{}^\au\nn\\
&&-\frac{1}{2}(\x_\ta\ki^\tb)\xx_{\a\tb}{}^\au
+\frac{1}{2}(\x_\ta\xx_\tb{}^\au)\ki^\tb_\a
+\x_{\a\tb}(\ki_\ta\xx^{\tb\au})
-\frac{1}{2}\x_{\a\ta}(\ki_\tb\xx^{\tb\au})\\
\d^\wz_\x \xx^{\da\ta}{}^\au&=&
2\vi(\x_\tb\s^a\eps)^\da \hat{\cd}_a t^{[\tb\ta]\au}
-2(\x_\tc\xx_\tb{}^\av)\xx^{\da\ta}{}^\az \tt^{[\tc\tb]\au}\gg_{\av\az}
-\frac{1}{4}(\x_\tb\ki^\tb)\xx^{\da\ta\au}\\
\left(\d^\wz_\x v_m{}^\au\right)P^\ym{}_\au{}^\az
 & = & \vi L^{1/2}\x_\tc\s_m\xx^{\tc\az}
  \eea
The fields $U_a{}^\tb{}_\ta$ and the traceless part of
$\ki_a{}^\tb{}_\ta$ contain only quadratic terms in the spinors,
\bea
\widetilde{U}_a{}^\tb{}_\ta& =& \widetilde{\ki}_a{}^\tb{}_\ta\ =\
-\frac{\vi}{8}
\left(\widetilde{\ki^\tb\s_a\ki_\ta}
+2\gg_{\az\au}\widetilde{\xx_\ta{}^\az\s_a\xx^\tb{}^\au}\right)\\
U_a&=&\frac{\vi}{8}\left(\ki^\ta\s_a\ki_\ta
+2\gg_{\az\au}\xx_\ta{}^\az\s_a\xx^\ta{}^\au\right)
\eea
while the trace part of $\ki_a{}^\tb{}_\ta$, corresponding to the
$U(1)$ part of the
initial $U(4)$ connection, contains the fieldstrengths of the
antisymmetric tensor
\bea
\ki_a&=&\vi H_a^*L^{-1}
-\frac{3\vi}{8}\left(\ki^\ta\s_a\ki_\ta
+2\gg_{\az\au}\xx_\ta{}^\az\s_a\xx^\ta{}^\au\right)
\eea

These expressions of the supersymmetry transformations
can be compared to the component level
results \cite{Cha81a}, \cite{dR85}, they are written automatically in terms of
supercovariant fieldstrengths.

\section{The $\frac{SO(6,n)}{SO(6)\times SO(n)}$ sigma model}
\label{sigma}

All we know about scalars $\tt^{[\tb\ta]}{}^\au$ at dim 0 is that they satisfy
the relations \equ{th}, implied by the Binachi Identities for the 3--form.
The general solution of this equation can be written in a form
\be
\tt^{[\tb\ta]}{}^\au\ =\ \tt^\oo{}^{[\tb\ta]}{}^\az \cg_\az{}^\au
\ee
where $\tt^\oo{}^{[\tb\ta]}{}^\az$ is a particular solution and $\cg$ is
an element of the Lie group leaving the metric $\gg$ invariant. Since the
signature of this metric was fixed to (6,n), this means that $g$ is an element
of $SO(6,n)$.

There is a particular solution of \equ{th} which is already known and
has a special meaning:
\be
\tt^\oo{}^{[\tb\ta]}{}^\az\ =\ \left(\tt^\oo{}^{[\tb\ta]}{}^z,\quad 0\right)
\ee
where the
central charge indices $\az$ were splitted in two groups,
$\az=(z,\bar{z})$, with $z=1..6$ and $\bar{z}=1..n$, and 
$\tt^\oo{}^{[\tb\ta]}{}^z$ are the covariantly constant $\tt$ and $\hh$
matrices used in the pure N-T supergravity case \cite{GK02}.
In a similar way, one defines the matrix of constants
\be
\qq^\oo_J{}^\az\ =\ \left(0,\quad \d_J^{\bar{z}}\right)
\ee
with $J=1..n$ and forms a $(6+n)\times(6+n)$ matrix as
\be
S^\oo\ =\ \left(\begin{array}{c}
\tt^\oo{}^{[\tb\ta]}{}^\az\\
\qq^\oo_J{}^\az
\end{array}\right).
\ee
Now one can verify that $S^\oo$ is an element of $SO(6,n)$ and obviously,
\be
S\ =\ S^\oo \cg\ =\ \left(\begin{array}{c}
\tt{}^{[\tb\ta]}{}^\az\\
\qq_J{}^\az
\end{array}\right),
\eqn{S}
corresponding to a general solution is a general element of
$SO(6,n)$. It is interesting to note that
the particular solution $S^\oo$ corresponds
to the ``uncoupled" sugra+6YM system.

At this stage multiplications
on the right by global elements of this
group $G=SO(6,n)$ are well-defined
\be
S\longrightarrow S\cg,\qquad \cg\in G.
\ee
The question is what subgroup $K$ of $G=SO(6,n)$ can act on the left
on $S$
\be
S\longrightarrow \ck^{-1} S,\qquad \ck\in K
\ee
such that the corresponding gauge transformation
is a symmetry of the theory.
On the one hand, since only the scalar components $\tt{}^{[\tb\ta]}{}^\az$ appear explicitly,
a local transformation which leaves invariant the upper $6$ rows of the 
matrix $S$ in \equ{S}
is a symmetry of the action. This is an $SO(n)$ rotation having the
representation
\be
\ck_{SO(n)}\ =\ \left(\begin{array}{cc}
\frac{1}{2}\d_{\td\tc}^{\tb\ta}&0\\
0&\ck_J{}^I
\end{array}\right).
\ee
On the other hand, the structure group of our superspace contains an $SU(4)$ factor
which is automatically implemented as local symmetry of the theory. In particular,
an $SU(4)$ transformation of a vector
\be
u^\ta \longrightarrow k^{-1}{}^\ta{}_\tb u^\tb
\ee
acts on the scalars $S$ with the representation
\be
\ck_{SU(4)}\ =\ \left(\begin{array}{cc}
k^\tb{}_{[\td}k^\ta{}_{\tc]}&0\\
0&\d_J{}^I
\end{array}\right).
\ee
In fact the constraint \equ{Dth} insures that
the $SU(4)$ connection $\hat{\F}^{\ta}{}_{\tb}$
is given as a function of the derivatives of the scalars,
\be
\left(\vd \tt^{[\tb\ta]\au}\right)\hh_{\au[\td\tc]}
\ =\ 2\d^{[\tb}_{[\td}\hat{\F}^{\ta]}{}_{\tc]}.
\ee

This concludes our identification of the
$\frac{SO(6,n)}{SO(6)\times SO(n)}$ sigma model parameterized by
the scalars $\tt^{[\tb\ta]}{}^\au$ subject to the relation
\equ{th}.

\end{appendix}

\bibliography{newREF}
\bibliographystyle{JHEP}

\end{document}